\documentclass[prd,twocolumn,preprintnumbers]{revtex4-1}

\usepackage[utf8]{inputenc}
\usepackage{listings}

\usepackage{amsmath}
\usepackage{amssymb}
\usepackage{graphicx}
\usepackage{booktabs,multirow,array}
\usepackage[table]{xcolor} 

\usepackage{amsmath}
\usepackage{epsfig}
\usepackage{graphicx}
\usepackage{color}
\usepackage[normalem]{ulem}
\usepackage{url}
\usepackage{float}
\usepackage[breaklinks, plainpages=false, colorlinks=true, anchorcolor=cyan, linkcolor=red, citecolor=cyan, urlcolor=magenta, bookmarks=false]{hyperref}

\usepackage[caption=false]{subfig}

\begin{document}
\renewcommand{\thefigure}{\arabic{figure}}
\setcounter{figure}{0}

 \def\I{{\rm i}}
 \def\E{{\rm e}}
 \def\D{{\rm d}}

\bibliographystyle{apsrev}

\title{Massive Black Hole Binaries and Where to Find Them with Dual Detector Networks}

\author{Kevin J. Shuman}
\email{kevinshuman@montana.edu}
\affiliation{eXtreme Gravity Institute, Department of Physics, Montana State University, Bozeman, Montana 59717, USA}

\author{Neil J. Cornish}
\email{ncornish@montana.edu}
\affiliation{eXtreme Gravity Institute, Department of Physics, Montana State University, Bozeman, Montana 59717, USA}

\begin{abstract}
    A single space-based gravitational wave detector will push the boundaries of astronomy and fundamental physics. Having a network of two or more detectors would significantly improve source localization. Here we consider how dual networks of space-based detectors would improve parameter estimation of massive black hole binaries. We consider two scenarios: a network comprised of the Laser Interferometer Space Antenna (LISA) and an additional LISA-like heliocentric detector (e.g.  Taiji);  and a network comprised of LISA with an an additional geocentric detector (e.g. TianQin). We use Markov chain Monte Carlo techniques and Fisher matrix estimates to explore the impact of a two detector network on sky localization and distance determination. The impact on other source parameters is also studied. With the addition of a Taiji or TianQin, we find orders of magnitude improvements in sky localization for the more massive MBHBs, while also seeing improvements for lower mass systems, and for other source parameters.
\end{abstract}

\maketitle

\section{Introduction}

The expansion of gravitational wave astronomy to the milli-Hertz band in the next decade is eagerly anticipated. Missions such as Laser Interferometer Space Antenna (LISA)~\cite{amaroseoane2017laser}, Taiji~\cite{taiji}, and TianQin~\cite{TianQin} will investigate the mysteries of black hole and galaxy formation~\cite{scilisaI}, explore the boundaries of general relativity~\cite{scilisaV,Gair2013TestingGR}, and probe many facets of cosmology~\cite{scilisaII,scilisaIII,scilisaIV,standardtaiji}, while potentially uncovering new phenomena and additional mysteries.

Most designs for future space-based gravitational wave detectors involve triangular constellations of spacecraft that will orbit the Sun while pirouetting about their own axes. This dance imprints upon the gravitational wave signal an impression that encodes the extrinsic source parameters - such as sky location and polarization. For LISA and Taiji, the result of this motion will allow massive black hole binaries (MBHBs) to be localized sufficiently well to perform interesting science~\cite{amaroseoane2017laser,scilisaI,scilisaV,scilisaII,scilisaIII,scilisaIV}. However, recent results~\cite{Blackholehunting} show that MBHBs with masses above $3\times 10^5 M_{\odot}$ exist so briefly in the LISA frequency band that achieving the sky localization needed to find electromagnetic counterparts will be challenging for those sources. It has also been shown that at low frequencies, where larger massive black holes signals reside, significant sky localization is hard to come by due to an eight-fold degeneracy in sky location that is only broken near the merger time~\cite{marsat2020exploring}. 

The goal of this paper is to investigate how adding a second LISA-like detector affects the parameter estimation for MBHBs, with a focus on sky localization and distance determination. The analysis is done for two scenarios: one with Taiji and another with TianQin. Our expectations are that by having a second detector, the combination of the time delay between the detectors, the different orientation of the antenna patterns, along with the combined effects of the detector motion will allow for better parameter estimation.

Dual LISA-like detector networks were first considered in Ref.~\cite{Crowder:2005nr}, but the study was limited to the inspiral signal, and Fisher information matrix estimates. More recently, similar studies have be performed for a LISA-Taiji network~\cite{ruan2019lisataiji,wang2021alternative} and a  LISA-TianQin network~\cite{Zhu:2021aat,wang2021forecast}. In this work we use a full insprial-merger-ringdown waveform model for systems with aligned spins (IMRPhenomD)~\cite{PhenomD1,PhenomD2} integrated with the full LISA, Taiji and TianQin responses~\cite{forward} to study the individual and dual detector parameter estimation capabilities. We compare results from a full Bayesian forecast using an MCMC algorithm with Fisher matrix parameter estimates. We find that the Fisher matrix estimates are very close to the MCMC results for dual detector networks, but can not be relied on for single detector analyses. We apply or analyses to two black hole population models, using simulated catalogs for light seed and heavy seed models~\cite{Barausse:2012fy,PhysRevD.93.024003}. We investigate the time evolving nature of the angular resolution focusing on six representative systems.

The paper is arranged as follows: In Section 2, we study and compare the posteriors given by the Fisher information matrix and the full Bayesian MCMC analysis. In section 3 we use the Fisher parameter estimation to analysis how different network configurations affect parameter estimation. In section 4 we ask when sources can be localized and look at the time evolving nature of the angular resolution. Lastly, in section 5 we discuss the results and conclude.

\section{Fisher and MCMC Comparison}

Ideally we would use full Bayesian forecasting based on MCMC exploration of the posterior distributions, but with hundreds of sources in the catalogs, and with ten different detector configurations to consider, the computational cost becomes prohibitive. A much cheaper alternative is to use the Fisher information matrix, which provides a Gaussian approximation to the likelihood that can be used to provide rapid forecasts of the parameter estimation capabilities of a detector. However, the Fisher matrix approximation breaks down when there are strong correlations between parameters, and/or for low signal-to-noise systems, and for situations where the priors are strongly curved, or alternatively, where the posterior distributions for some of the parameters cover a large fraction of the prior range~\cite{Vallisneri:2007ev}. What we find is that while the Fisher forecasts are unreliable for single detectors, they are very accurate for the dual detector configurations.

Below we discuss the background methodology, then describe how the comparison was made using a sample of 100 sources from the light and heavy seed catalogs.
To save computational cost, we restrict the MCMC-Fisher comparison to the LISA and LISA-Taiji network.

\subsection{Background Methodology}

Throughout this work we used IMRPhenomD waveforms and the LISA and TianQin detector responses as described by Cornish \& Shuman~\cite{Blackholehunting}, where the significant difference between LISA and TianQin are the arm lengths and orbits \cite{forward,tianqinorbits}. The IMRPhenomD waveforms emulate those of spin-aligned black hole binaries. The projected waveforms have eleven parameters: the masses, spins, luminosity distance, coalescence time and phase, sky location, polarization angle, and inclination angle. We produce parameter resolution estimates for all eleven parameters, but we are mostly focused on sky location and luminosity distance. For the Fisher estimates, we used the full waveform, while for the MCMC estimates we also considered how the parameters evolve with time.

The methodology for a Fisher parameter estimation is explained well in several places~\cite{lang,poisson}. The general idea is to approximate the likelihood distribution as a multivariate Gaussian distribution using the covariance matrix $C_{ij}$,

\begin{equation}
    p(s|\Tilde{\boldsymbol{\theta}}) \propto e^{-\delta\theta^{i}(C_{ij})^{-1}\delta\theta^{j}/2},
\end{equation}

\noindent where the indices run over the parameters and $\delta \theta^i$ is the difference between the true parameter value and the value that maximizes the probability. The inverse of the covariance matrix is the Fisher information matrix, $\Gamma_{ij}$, which we calculate using the amplitude and phase of the waveform, as opposed to the usual waveform calculation since the amplitude and phase are smooth functions, by

\begin{figure}[th!]

\includegraphics[width=10cm]{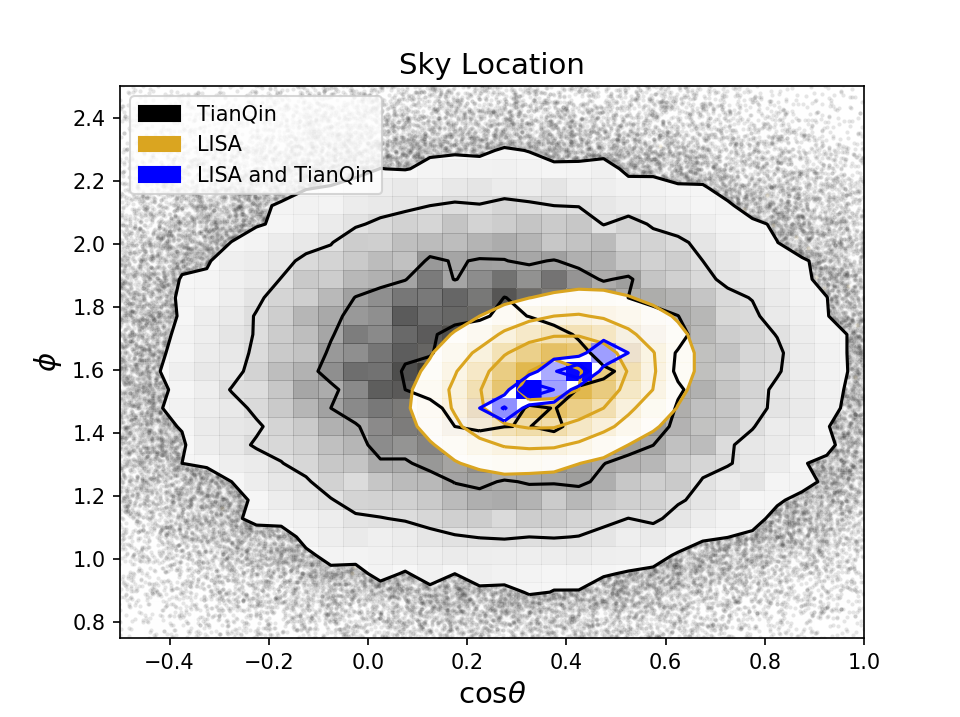}

\caption{
Sky localization 
derived from the Fisher matrix for the individual LISA and TianQin detectors, and for the dual network with both detectors. We see a significant decrease in sky area covered by the dual detector} \label{fig:FisherIll}
\end{figure}

\begin{equation}
    \Gamma_{ij} = 4 \int^{\infty}_{0} \frac{\frac{\partial A}{\partial \theta_i}\frac{\partial A}{\partial \theta_j} + A^2 \frac{\partial \Phi}{\partial \theta_i}\frac{\partial \Phi}{\partial \theta_j}}{S_{n}(f)} df,
\end{equation}

\noindent where $A$ is the waveform amplitude, $\Phi$ is the waveform phase, and $S_n(f)$ is the noise power spectrum, which is given in \cite{Blackholehunting}. Having two constellations results in the total posterior distribution being the product of the posterior distributions for each constellation. This leads to us defining the total Fisher matrix as the sum of the Fisher matrix for each constellation, $\Gamma^{(Tot)}_{ij} = \Gamma^{(1)}_{ij} + \Gamma^{(2)}_{ij}$.

Since the approximate shape of our distribution is a Gaussian, it is typical to use the area of the error ellipse, $\Delta \Omega$, as the measure of the angular resolution of our detector, which tells us the probability is $e^{-1}$ for a source to lie outside the error ellipse. The expression is given by
\begin{equation}
    \Delta \Omega = 2\pi \sqrt{C^{(Tot)}_{\mu \mu} C^{(Tot)}_{\phi \phi} - (C^{(Tot)}_{\mu \phi})^2},
\end{equation}
where $\mu = \cos(\theta)$, $\theta$ the polar angle, and $\phi$ is the azimuthal angle. Figure~\ref{fig:FisherIll} compares the error ellipses for the individual LISA and TianQin detectors to the dual detector case. Combing the information from the two detectors significantly improves the predicted sky localization.

It was found that this approach of using the error ellipse was not well suited for the single detector analysis. The uncertainties tend to be rather large and do not take into account the priors. Instead, we used our calculation for the total Fisher matrix to generate a Gaussian distribution and sampled from that distribution, excluding points that ventured outside the priors. Once we had a distribution, we could find the angular resolutions by choosing a confidence interval and finding the area. This method is slower, but generally safer and more accurate and was used instead of using the error ellipse. As for determining the uncertainty in the luminosity distance, we also used this distribution to calculate a confidence interval.

A single source gives us a single angular resolution associated with that source, so to do a meaningful study of how a second detector affects sky localization, we need to consider many representative systems. Here we use two catalogs of sources from the LISA Data Challenge~\cite{babak_petiteau}. These populations and their properties are outlined in Refs.~\cite{Barausse:2012fy,PhysRevD.93.024003,scilisaI}. The two catalogs cover distinct mass ranges, corresponding to light and heavy seed scenarios. The light seed (popIII) catalog holds sources with total masses mostly ranging from $10^{3} - 10^{6} M_{\odot}$ with median $\sim10^{3} M_{\odot}$. While the second heavy seed catalog (Q3\_nodelay), contains sources with total masses mostly ranging from $10^{4} - 10^{8} M_{\odot}$ with median $\sim10^{6} M_{\odot}$. From here on, we will refer to the popIII sources as the light seeds and the Q3\_nodelay sources and heavy seeds. It is worth noting that the catalogs are of populations of massive black hole binaries that were formed from specific seeds, which implies the sources have predominately large redshifts.

\subsection{Comparison}
We used a Fisher parameter estimation to make the computational time feasible, but it is an approximation whose assumption of a Gaussian posterior could be false. For a more realistic posterior, we adopted and modified the replica exchange (parallel tempering) Markov chain Monte Carlo code used in \cite{Blackholehunting} to find the LISA, TianQin, and dual detectors posteriors for six sources (Table III) to observe the characteristics of the posteriors before doing the full 100 source comparison. We used the $40^{\circ}$ ahead and flipped (described later on) configuration, which showed the most promising results in \cite{ruan2019lisataiji}. 

\begin{figure*}[th!]
\subfloat{\includegraphics[height=7cm,width=.49\linewidth]{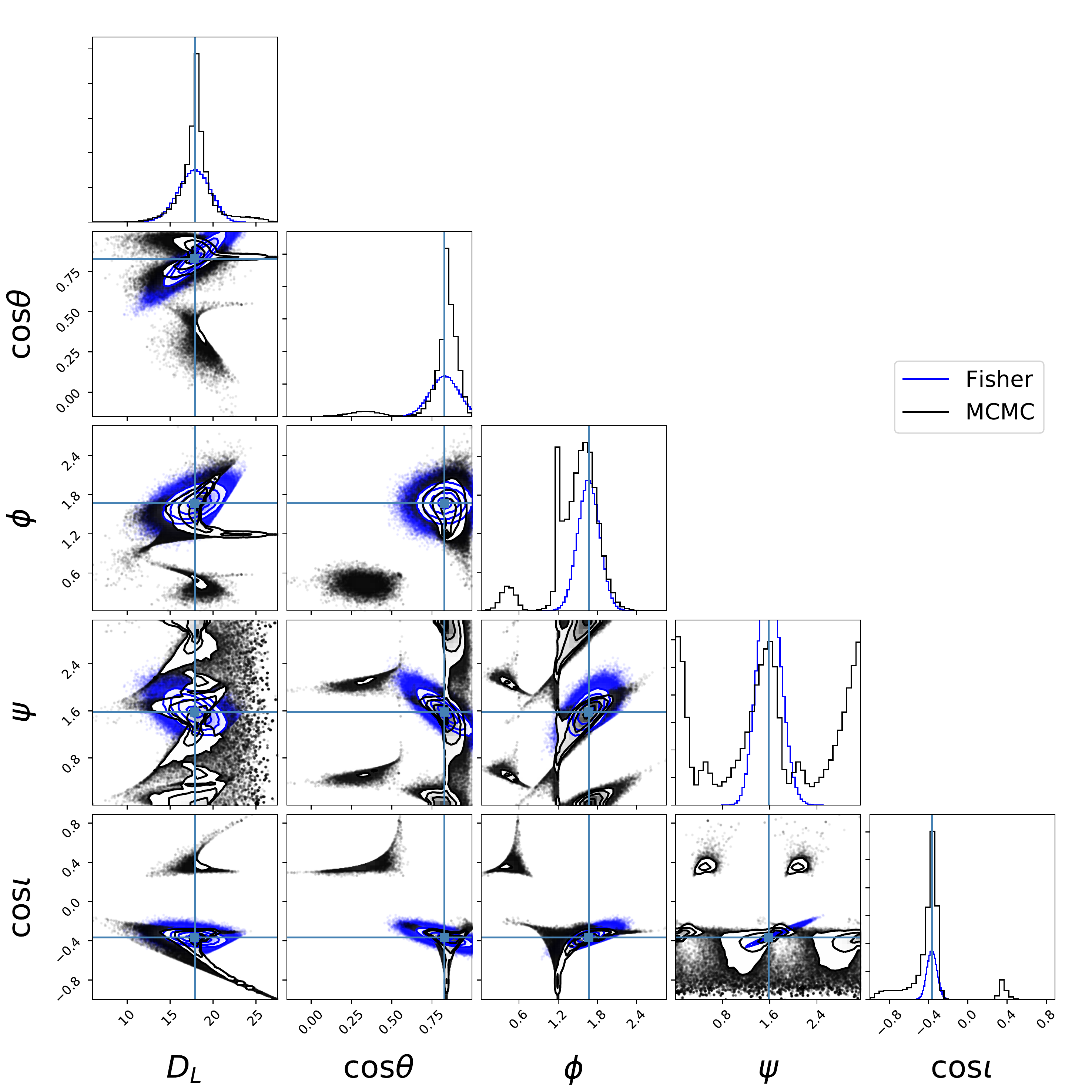}\label{fig:ex}}
\hspace*{\fill}
\subfloat{
\includegraphics[height=7cm,width=.49\linewidth]{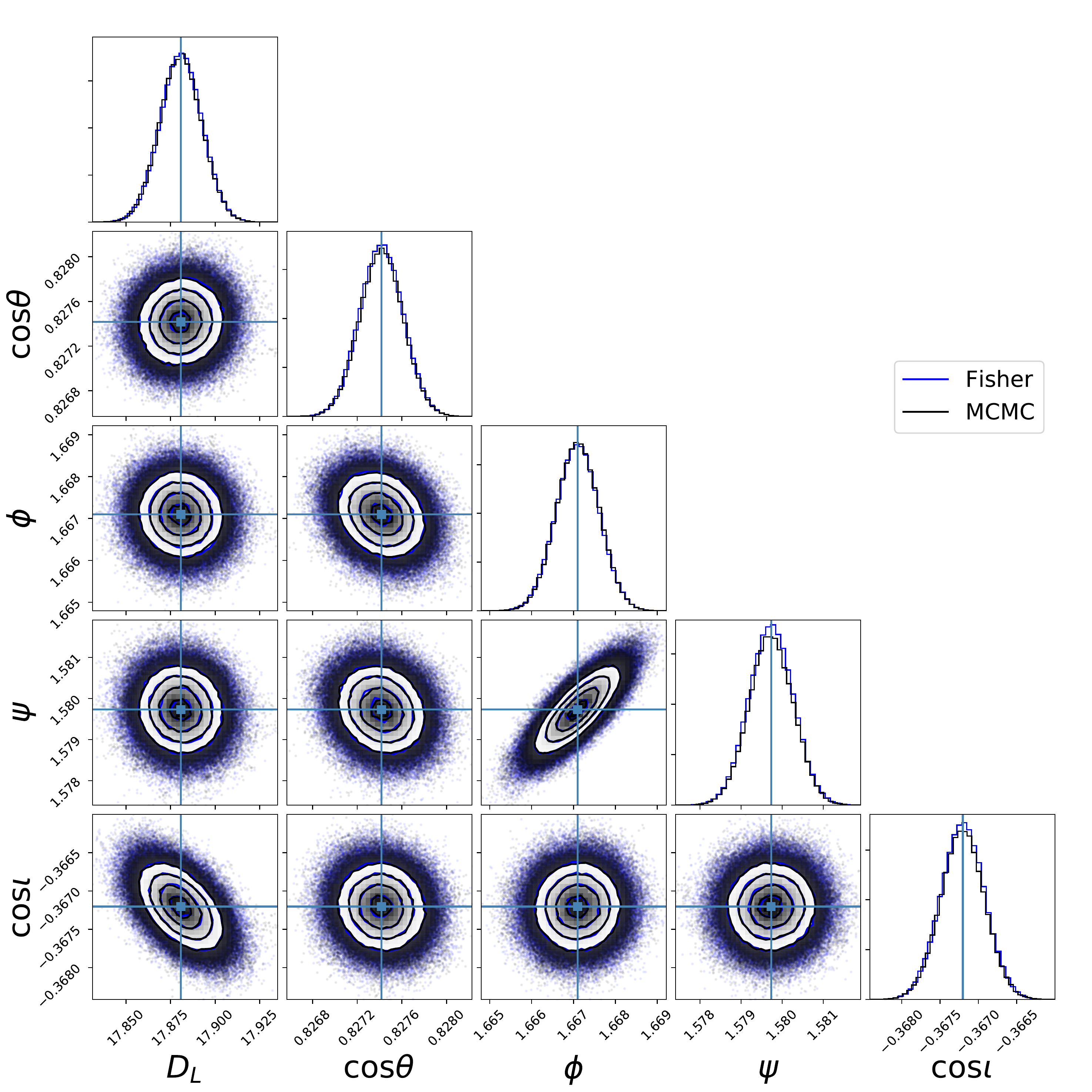}\label{fig:dual_ex}}

\caption{Extrinsic parameter corner plots of source 6 with the single LISA (left) and LISA-Taiji (right) configurations. The blue vertical and horizontal lines represent the true values of the parameters. The blue distribution is a Gaussian distribution with covariance the inverse of the Fisher matrix.} \label{fig:LISA_Corner}
\end{figure*}

\begin{figure*}[th!] 
\subfloat{
\includegraphics[height=7cm,width=.49\linewidth]{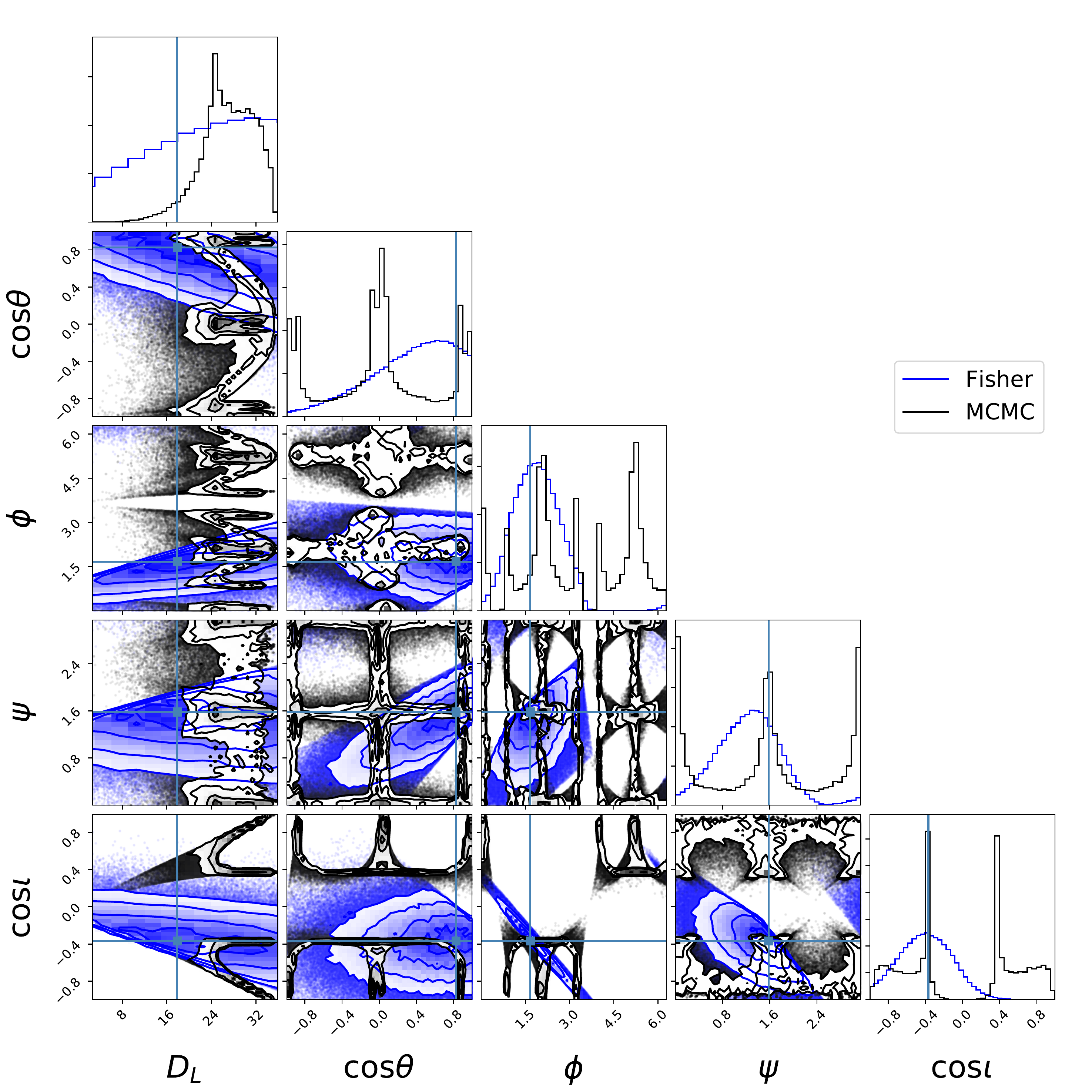}\label{fig:TianQin_ex}}
\hspace*{\fill}
\subfloat{\includegraphics[height=7cm,width=.49\linewidth]{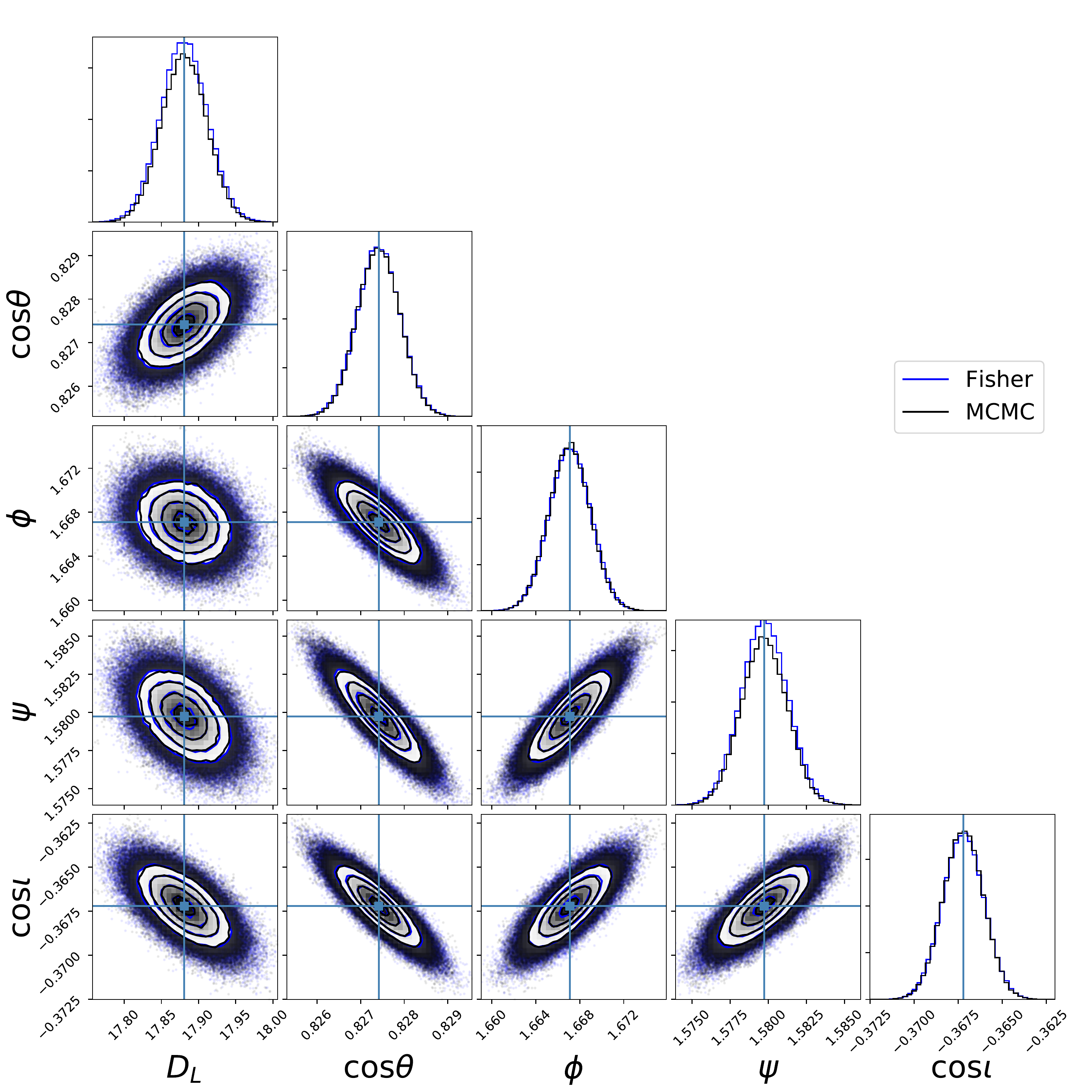}\label{fig:TianQin_dual_ex}}

\caption{Extrinsic parameter corner plots of source 6 with the TianQin (left) and LISA-TianQin (right) configurations. The blue vertical and horizontal lines represent the true values of the parameters. The blue distribution is a Gaussian distribution with covariance the inverse of the Fisher matrix.} \label{fig:TianQin_Corner}
\end{figure*}

Figure~\ref{fig:LISA_Corner} and Figure~\ref{fig:TianQin_Corner} show the marginalized posteriors for the extrinsic parameters for combinations of LISA, a second LISA, and TianQin using source 6 (one of the more massive sources). In Figure~\ref{fig:LISA_Corner} we see that the posterior for the $\phi$ and $\cos{\theta}$ has multiple maxima while that for the LISA-Taiji case does not and is fitted well by a Gaussian distribution. This pattern was true for all six of the sources. The single maximum and the Gaussian shape suggests that the Fisher parameter estimation is valid as are the values for the angular resolutions we calculated before. Similarly, Figure~\ref{fig:TianQin_Corner} shows that TianQin has degeneracies in sky location and that the LISA-TianQin case does not and can be fitted with a Gaussian distribution. From these results we can see that the posterior for the Fisher sampling posterior are sufficient for the dual detector case. Thus we restricted ourselves to comparing the single detector posteriors.

We wanted to consider how well the Fisher parameter estimation does in comparison to the MCMC for the single detector case. To do this, we chose 100 sources randomly with signal-to-noise above 13 from the two catalogs. We found the 50\% and 90\% confidence regions of the sky location marginalized posteriors for these sources using both the Gaussian and MCMC posteriors distributions. The best case scenario would be that the angular resolutions we obtain would line along the line of slope 1, suggestions the angular resolutions of the Fisher sampling method are equivalent to the MCMC method. In such a case we would not need to run a full MCMC parameter estimation, and we could get away with the quicker Fisher sampling method.

\begin{figure*}[th!]
\subfloat{\includegraphics[height=7cm,width=.49\linewidth]{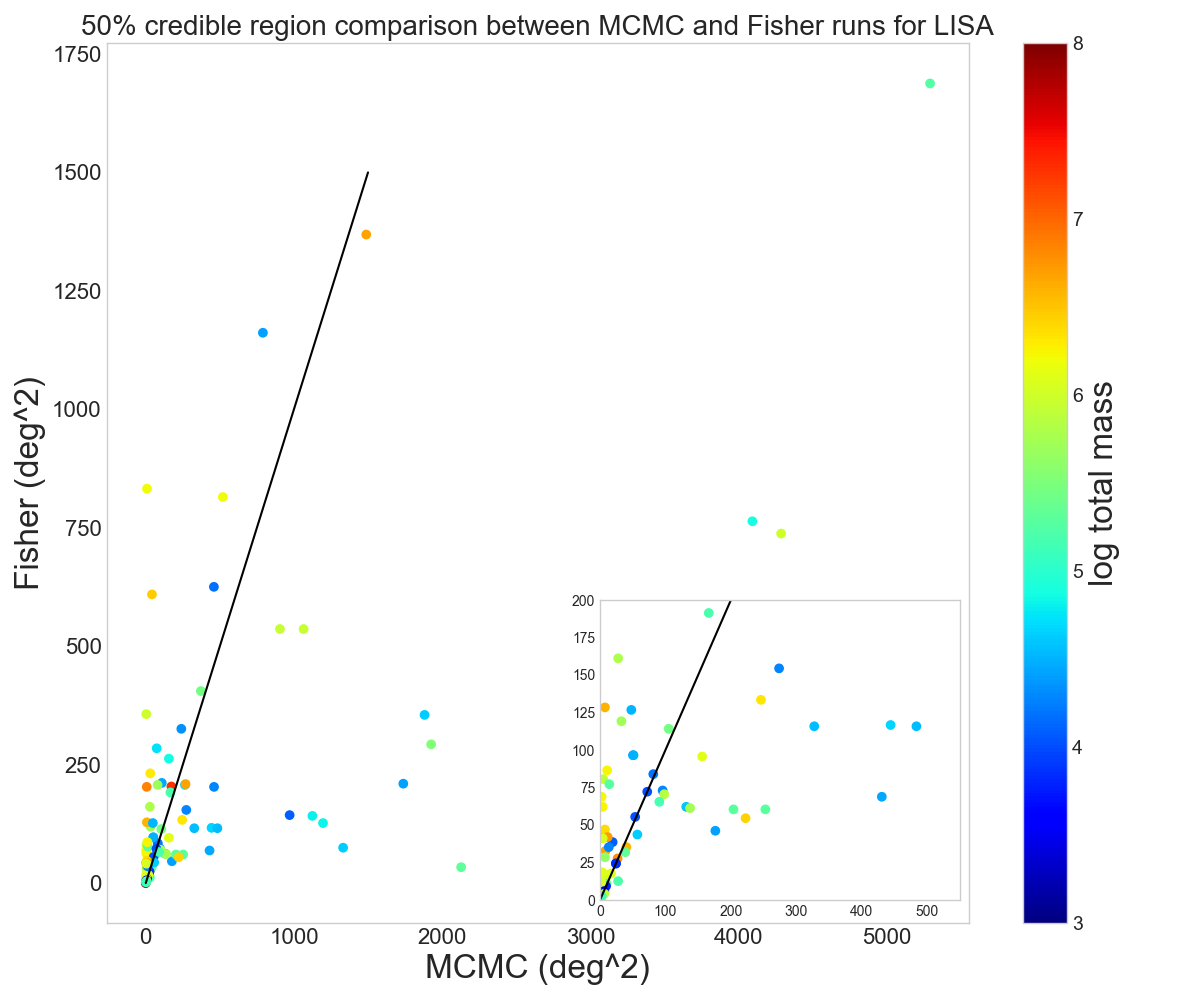}\label{fig:Q3_nodelayA}}
\hspace*{\fill}
\subfloat{\includegraphics[height=7cm,width=.49\linewidth]{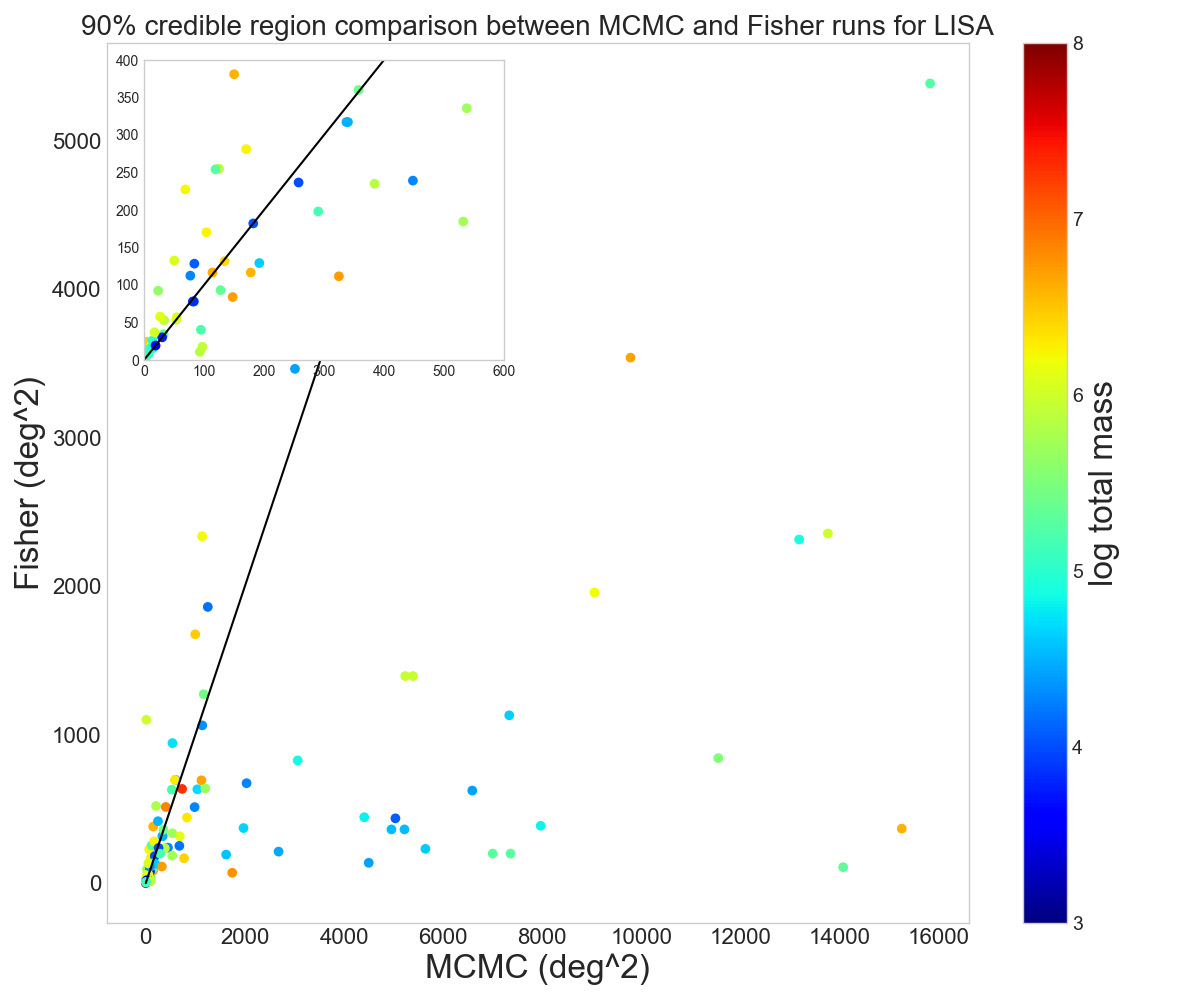}\label{fig:Q3_nodelayB}}

\medskip
\subfloat{\includegraphics[height=7cm,width=.49\linewidth]{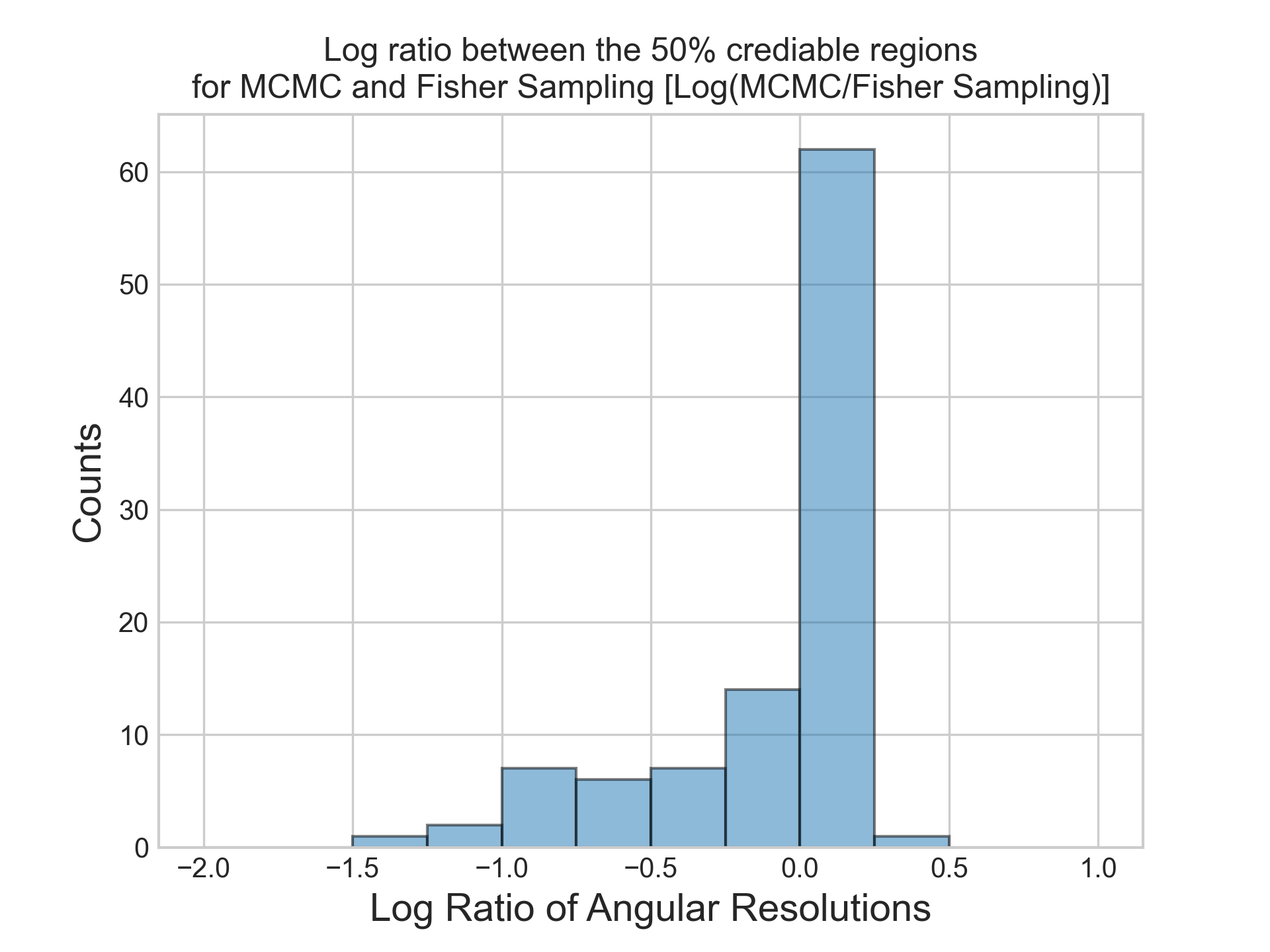}\label{fig:Q3_nodelayC}}
\hspace*{\fill}
\subfloat{\includegraphics[height=7cm,width=.49\linewidth]{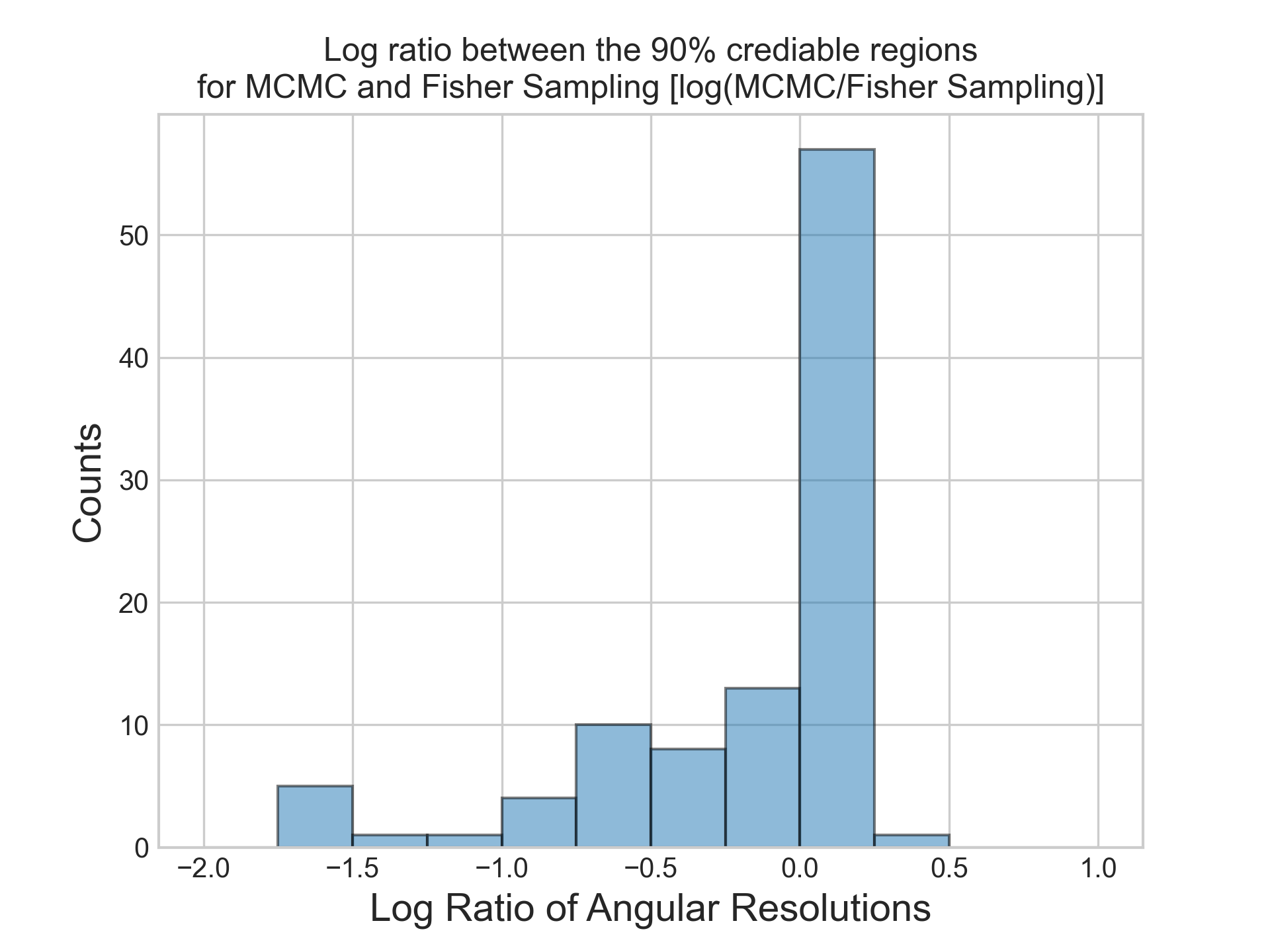}\label{fig:Q3_nodelayD}}

\caption{The top two scatter plots show the 50\% and 90\% confident regions between the MCMC method and the Fisher matrix for sky location posteriors for 99 of the 100 sources (one source was a considerable outlier that disturbs the plot with coordinates (2156, 3787) and (15520, 11991) for the 50\% and 90\% confident regions). The black line shows the ideal situation where the MCMC method gives the same result as the Fisher matrix method. the bottom two histograms show the log ratio of the MCMC angular resolution and the Fisher matrix angular resolution for the two confidence regions.} \label{fig:MCMC_Log_Ratio}
\end{figure*}

Figure~\ref{fig:MCMC_Log_Ratio} gives an idea how the Fisher sampling compares to the MCMC. In the scatter plot shown we see substantial deviations from the best case of the angular resolutions being equal, suggesting it is generally a bad idea to use a Fisher parameter estimation for a single detector analysis. Also note that there seems to be no relationship between the total source mass and the difference in angular resolution. The histograms of the log ratios give a more quantitative picture. We see that the majority of the sources are roughly a factor of 2-3 off, suggesting the Fisher parameter estimation is picking up one of the modes of a bi-modal posteriors, but we see $\sim$40\% of the sources have a greater difference, with the Fisher result being the larger of the two, telling us we can't assume the resulting posteriors are bi-modal having done a Fisher parameter estimation and that the modes of the MCMC results are poorly approximated by a Gaussian distribution. While we are only considering the sky location here, given the corner plots of the single detector cases, it seems safe to suggest the trend carries over to the other parameters as well.

From these results we obtained a picture of how we would proceed with the rest of our analysis. We gathered that we cannot use a Fisher parameter estimation for the single detector case. On the other hand, we found that the Fisher parameter estimation works very well for the dual detector case. In the next section we used the near equivalence of the Fisher sampling method in the dual detector case to determine the angular resolution and luminosity distance uncertainty of different dual network configurations.

\section{Fisher Analysis}

To motivate why we believe the advantages of adding a second detector are worthwhile consider Figure ~\ref{fig:Fisher_Ill}. Here we have three Gaussian distributions generated using the sampling method discussed above. On the left and in middle, we have sky location marginalized posterior distributions for a source generated using the LISA and TianQin responses, respectively. On the right, we add the Fisher matrices of the other two distributions to get a new posterior distribution and marginalize to get the sky location distribution. The axes here are fixed to show the relative scale of the changes. We see a significant decrease in area for the dual detector case. This result is not source specific but rather comes from the additional information gained when a second detector is added. The Fisher matrix is a measure of the information about the uncertainties carried by the a set of source parameters as a result of the waveform and response function. The two Fisher matrices for the two detectors carry information about how changes in parameters about the original source parameters affect their waveform measurements. In adding the matrices together we reduce the uncertainties because we reduce the size of the changes that would produce a similar waveform because each detector allows for different sized changes for different parameters and has stronger or weaker correlations among those parameters, which in the end all need to be reduced to a compromise in what changes are allowed that produce a similar waveform.

In the case of the LISA-TianQin network, the detectors are fixed, but for one of the two LISA-like constellations (e.g. LISA and Taiji) we have the freedom to place the detector. Here we considered seven configurations for the LISA-LISA type network. For convenience later, we will give the different configurations names, which will be parenthesized next to the configuration. The choices were influenced by LISA's planned position with respect to the Earth and somewhat due to the potential costs of placing a second detector. The first configuration sits at the same location as LISA but is flipped (LLF). That is, LISA will be tilted from the plane of the ecliptic by either $60^{\circ}$ or $-60^{\circ}$ (the idea of combining regular and flipped arrays was first introduced at the Tenth International LISA Symposium by one of the authors~\cite{LISAX}). The term flipped refers to the angle different to LISA. In our study, the LISA detector was set to $60^{\circ}$ and the flipped detector at $-60^{\circ}$. The idea behind the flipped configuration is that the antenna patterns are then mis-aligned, which allows for a better measurement of the polarization and inclination of the source, and therby breaks degeneracies in the sky and distance localization. The other configurations are: $10^{\circ}$ ahead in orbit (LL10), $10^{\circ}$ ahead in orbit and flipped (LLF10), $30^{\circ}$ ahead in orbit (LL30), $30^{\circ}$ ahead in orbit and flipped (LLF30), $40^{\circ}$ ahead in orbit (LL40), $40^{\circ}$ ahead in orbit and flipped (LLF40). We also consider LISA and TianQin (LT) and LISA. The idea behind the orbital separation is to allow for triangulation via time of arrival information (like for the terrestrial LIGO/Virgo network). The orbital separation can be combined with the orientation flip, but the added benefit of the flip diminishes as the orbital separation grows since the antenna patterns naturally become misaligned for large orbital separations even without the flip.

\begin{figure}[th!]
\subfloat{\includegraphics[height=4cm,width=.45\textwidth]{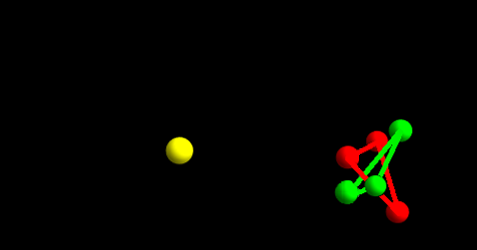}}

\subfloat{\includegraphics[height=4cm,width=.45\textwidth]{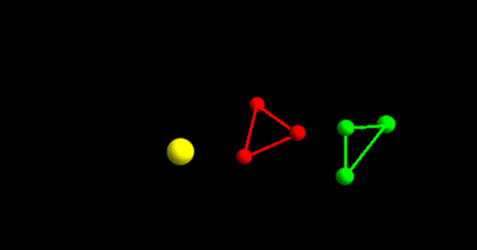}}

\caption{Illustrations of the LLF and LL40 configurations.} \label{fig:Config}
\end{figure}

\begin{table*}[th!]
\centering
\resizebox{\textwidth}{!}{
\begin{tabular}{@{} l*{7}{>{$}c<{$}} @{}}
\toprule
Configuration & \multicolumn{3}{c@{}}{Light Seeds} & \multicolumn{3}{c@{}}{Heavy Seeds} \\
\cmidrule(l){2-4}
\cmidrule(l){5-7}
& Mean \pm STD & Median & \textnormal{\% with }\Delta \Omega < 10 {\rm deg}^2 & Mean \pm STD & Median & \textnormal{\% with } \Delta \Omega < 10 {\rm deg}^2 \\
\midrule
\multirow{1}{*}{LISA} &  1130 \pm  1640  &  610  &  0.6  &  1020 \pm  2230  &  277  &  7.46 \\
\multirow{1}{*}{TianQin} &  4190 \pm  3830  &  2920  &  0.24  &  4310 \pm  5780  &  1070  &  7.94 \\
\multirow{1}{*}{LLF} &  477 \pm  810  &  250  &  3.36  &  3.65 \pm  6.49  &  0.825  &  87.3 \\
\multirow{1}{*}{LL10} &  363 \pm  507  &  200  &  3.12  &  6.38 \pm  8.75  &  2.29  &  79 \\
\multirow{1}{*}{LLF10} &  324 \pm  475  &  179  &  4.68  &  2.72 \pm  5.6  &  0.52  &  91.1 \\
\multirow{1}{*}{LL30} &  157 \pm  229  &  83.9  &  6.36  &  2.32 \pm  5.1  &  0.395  &  93.2 \\
\multirow{1}{*}{LLF30} &  149 \pm  219  &  77.2  &  7.44  &  1.58 \pm  4.18  &  0.218  &  95.9 \\
\multirow{1}{*}{LL40} &  120 \pm  178  &  62.1  &  7.68  &  1.74 \pm  4.28  &  0.26  &  95.6 \\
\multirow{1}{*}{LLF40} &  113 \pm  165  &  57.1  &  9.24  &  1.32 \pm  3.78  &  0.168  &  96.7 \\
\multirow{1}{*}{LT} &  105 \pm  148  &  60.4  &  6.84  &  2.26 \pm  5.48  &  0.321  &  93.7 \\
\bottomrule
\end{tabular}}
\caption{\label{Table_II}The angular resolution results of the various configurations between the two catalogs, light seeds and heavy seeds. The results here are given by calculating the 90\% confident region of the Fisher sampling distriution. Shown here are the mean, standard deviation, and median of the angular resolutions and the number of sources with angular resolution less than 10 square degrees, where there were 833 light seeds and 630 heavy seeds. }
\end{table*}

\begin{figure}[th!]
\subfloat{\includegraphics[height=7cm,width=.48\textwidth]{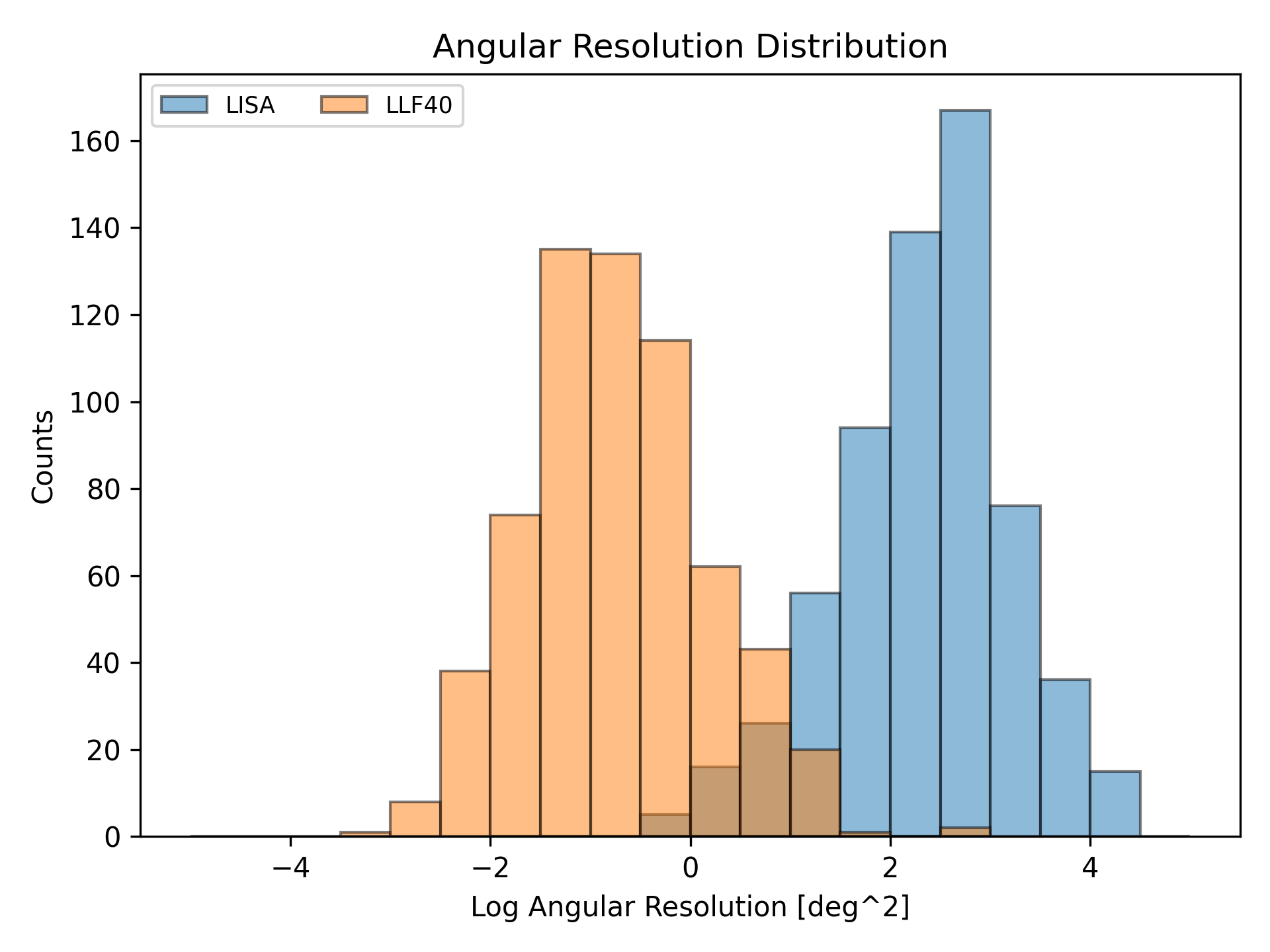}}

\caption{Here are two distributions of angular resolutions, $\Delta \Omega$, for the heavy seeds for LISA and of the LLF40 configuration.} \label{fig:Ang_Res_Dist}
\end{figure}

\begin{table}[th!]
\centering
\resizebox{\linewidth}{!}{
\begin{tabular}{@{} l*{5}{>{$}c<{$}} @{}}
\toprule
Configuration & \multicolumn{2}{c@{}}{Light Seeds} & \multicolumn{2}{c@{}}{Heavy Seeds} \\
\cmidrule(l){2-3}
\cmidrule(l){4-5}
& Mean \pm STD & Median & Mean \pm STD & Median \\
\midrule
\multirow{1}{*}{LISA} &  0.274 \pm  0.18  &  0.242  &  0.225 \pm  1.23  &  0.121 \\
\multirow{1}{*}{TianQin} &  0.434 \pm  0.237  &  0.401  &  0.333 \pm  0.224  &  0.29 \\
\multirow{1}{*}{LLF} &  0.215 \pm  0.155  &  0.191  &  0.0194 \pm  0.035  &  0.00697 \\
\multirow{1}{*}{LL10} &  0.229 \pm  0.16  &  0.203  &  0.0448 \pm  0.0656  &  0.0195 \\
\multirow{1}{*}{LLF10} &  0.215 \pm  0.155  &  0.193  &  0.0194 \pm  0.0351  &  0.00695 \\
\multirow{1}{*}{LL30} &  0.224 \pm  0.159  &  0.196  &  0.0265 \pm  0.0442  &  0.0101 \\
\multirow{1}{*}{LLF30} &  0.215 \pm  0.154  &  0.193  &  0.0196 \pm  0.0353  &  0.00699 \\
\multirow{1}{*}{LL40} &  0.222 \pm  0.158  &  0.196  &  0.0237 \pm  0.0399  &  0.00864 \\
\multirow{1}{*}{LLF40} &  0.215 \pm  0.154  &  0.194  &  0.0198 \pm  0.0355  &  0.007 \\
\multirow{1}{*}{LT} &  0.251 \pm  0.172  &  0.224  &  0.033 \pm  0.0547  &  0.0123 \\
\bottomrule
\end{tabular}}
\caption{\label{Table_III} The uncertainty in the luminosity distance in gigaparsecs for the various configurations between light and heavy seeds. The results here are given by calculating the 90\% confidence interval of the luminosity distance using the Fisher sampling distribution.}
\end{table}

LISA and TianQin have degeneracies in sky location (see Figure~\ref{fig:LISA_Corner} and Figure~\ref{fig:TianQin_Corner}), which implies those results of the Fisher parameter estimations are optimistic. Regardless, in Table I we see improvements in sky localization for the light seeds and orders of magnitude better sky localization within the heavy seeds. This difference between the two catalogs is expected. Lower mass sources stay within the LISA frequency band for a lot longer than higher mass sources. The overall higher angular resolution for the smaller sources deals with the high redshifts and low masses producing lower SNR signals. The advantage we utilize is that the second detector will sample a distinct portion of the detector response, which will allow for a better estimate of the extrinsic parameters. But if the source is in the LISA frequency band for long enough, we already have a significant portion of the response sampled, as if we have a stationary detectors at different places along the orbits. For this reason, we see that as we separate the LISA detectors further apart the angular resolution becomes smaller in all cases.

In a similar manner, Table II shows the mean, standard deviation, and median of the luminosity distance in gigaparsecs. These results were calculated by determining the 90\% confidence region of the marginalized Fisher sampling distribution for each of the sources of the sources catalogs in each configuration. We see there is little effect on the luminosity distance for the light seeds. On the other hand, we see an order of magnitude improvement from the single detectors for all the flipped dual networks with the heavy seeds. Keep in mind that the single detector posteriors are not Gaussian and the results for the single detectors should be considered ballpark estimates.

\section{Posterior Time Evolution}

\begin{table*}[tp]
\centering
\resizebox{\textwidth}{!}{
\begin{tabular}{cccccccc}
\toprule
Parameter & Symbol & Source 1 & Source 2 & Source 3 & Source 4 & Source 5 & Source 6 \\
\midrule
Mass 1 ($M_{\odot}$) & $M_1$ & 4.626772e+04 & 1.578363e+06 & 1.418998e+05 & 2.815416e+06 & 4.521938e+05 & 1.564456e+07 \\

Mass 2 ($M_{\odot}$) & $M_2$ & 6.315933e+03 & 1.304587e+06 & 2.398065e+04 & 1.080808e+06 & 9.299166e+04 & 3.489898e+06 \\

Spin 1 & $\eta_1$ & 9.880200e-01 & 7.069900e-01 & 4.087800e-01 & 5.489800e-01 & 3.175600e-01 & 9.628800e-01 \\

Spin 2 & $\eta_2$ & 6.512500e-01 & 4.800900e-01 & 7.000000e-01 & 9.395300e-01 & 4.745300e-01 & 9.880200e-01 \\

Coalescence Phase & $\phi_1$ & 2.552669e-01 & 2.314101e-01 & 3.140781e+00 & 1.516281e+00 & 2.797483e+00 & 1.669934e+00 \\

Coalescence Time (s) & $t_c$ & 2.179900e+07 & 2.827100e+07 & 2.991200e+07 & 3.010700e+07 & 3.157800e+07 & 3.106000e+07 \\

Luminosity Distance (Gigaparsec) & $D_L$ & 7.789894e+01 & 1.649478e+01 & 1.288355e+01 & 3.996195e+01 & 1.438562e+01 & 1.788077e+01 \\

Sine of Ecliptic Latitude  & $\sin{\theta}$ & 3.700845e-01 & 8.596543e-01 & 1.148387e-01 & 7.028475e-01 & 3.684582e-01 & 8.274171e-01 \\

Ecliptic Longitude & $\phi$ & 1.562200e+00 & 5.979600e+00 & 1.851500e+00 & 3.761400e+00 & 7.937100e-01 & 1.667100e+00 \\

Polarization Angle & $\psi$ & 3.833270e-01 & 1.924025e+00 & 5.117965e-01 & 2.114386e-01 & 2.381472e+00 & 1.579731e+00 \\

Cosine of Inclination Angle & $\cos{\iota}$ & 9.763423e-01 & 2.920247e-02 & -4.304167e-01 & -4.332995e-01 & -8.507850e-01 & -3.672035e-01 \\
\bottomrule
\end{tabular}}

\caption{Source parameters for the six sources used in the MCMC parameter estimation}
\end{table*}

Localizing a source after the black holes merge might be useful to some, but to the physics community we would hope to find the sources prior to merger. So we wanted to ask when can we localize the source. To do this, we considered six of the 100 sources (Table III) and found how their posteriors evolved with time by considering the signal up to merger, up to a day before merger, and up to six days before merger. 

\begin{figure*}[h]
\subfloat{\includegraphics[height=7cm,width=.49\linewidth]{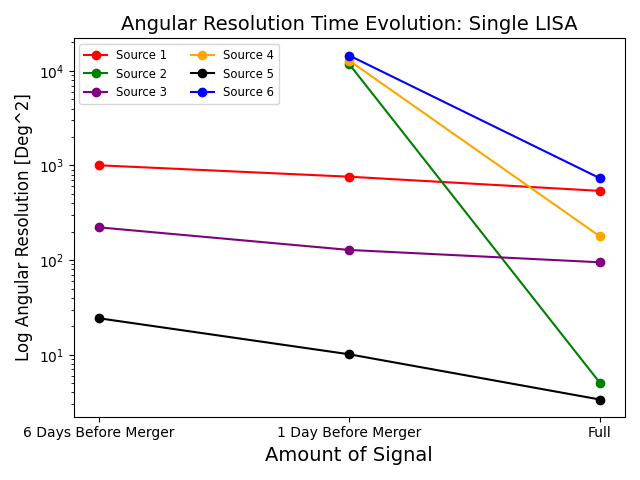}\label{fig:LISA_Ang}}
\hspace*{\fill}
\subfloat{\includegraphics[height=7cm,width=.49\linewidth]{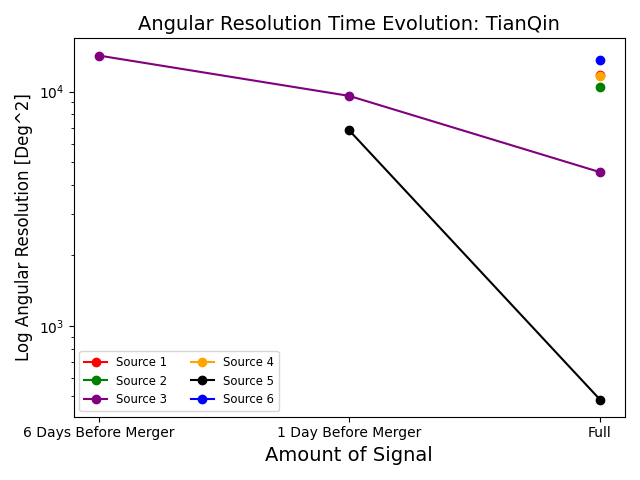}\label{fig:TianQin_Ang}}

\subfloat{\includegraphics[height=7cm,width=.49\linewidth]{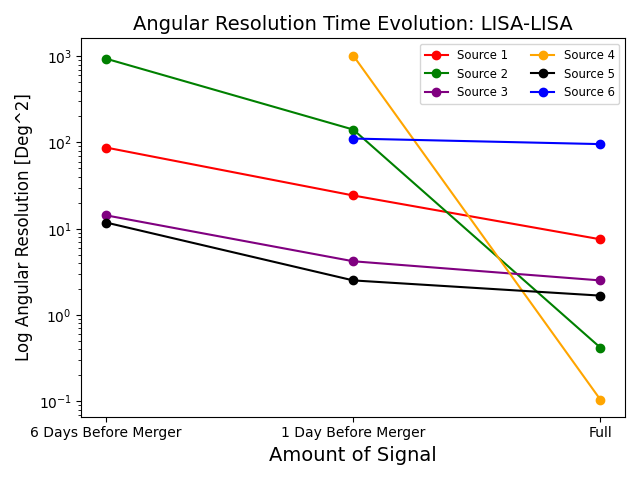}\label{fig:LISA_LISA_Ang}}
\hspace*{\fill}
\subfloat{\includegraphics[height=7cm,width=.49\linewidth]{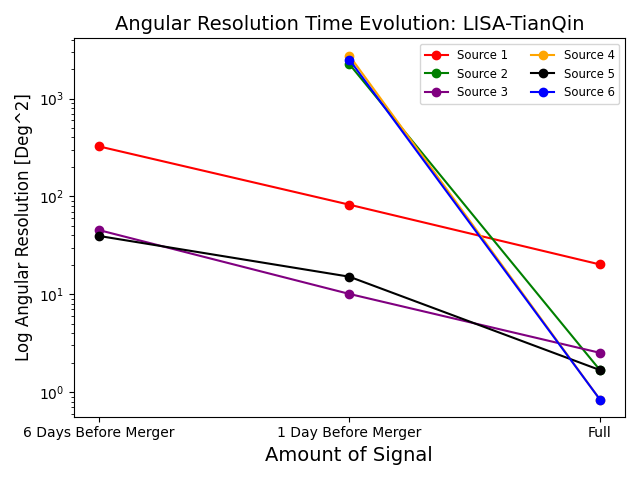}\label{fig:LISA_TianQin_Ang}}

\caption{The plots show the time evolving angular resolution, $\Delta \Omega$, of the six chosen sources with the single detectors and the dual detectors. Points missing from the plots imply the sources were not found by the MCMC.} \label{fig:Time_Ang}
\end{figure*}

Figure~\ref{fig:Time_Ang} shows the time evolving angular resolutions of the six sources for LISA, LISA-Taiji, TianQin, and LISA-TianQin using the LLF40 configuration. The angular resolutions for these sources were calculated using the 90\% credible interval of the marginalized posterior distributions. In all detector networks we miss sources (sources not being detected) six days before merger. For the dual detector cases, we manage to find at least three of the six sources with source 1 and 3 having better angular resolution than those sources with the full signal in the single detectors six days before merger. We also see that the shared sources between the single detectors and the dual networks are improved pre-merger.

\section{Discussion}

Much of the science space-based gravitational wave detectors do relies on the ability for these detectors to localize gravitational wave sources. In this paper, we have shown that having a second detector would substantially improve sky localization and significantly improve luminosity distance uncertainty for more massive MBHBs, up to a few orders of magnitude for sky location and up to an order of magnitude for luminosity distance uncertainty. The addition of a second, well-placed LISA-like detector makes it possible to localize the majority of detectable MBHB sources. The posterior distributions also become more accurately approximated by a Gaussian distribution. These advantages will do two things: first make parameter estimation significantly better and second improve the speed of our analysis. We can substitute determining the posterior distribution using an MCMC by using a Gaussian distribution determined from the Fisher matrix without loss of accuracy. 

And with the addition of a second detector we found that we can detect more sources prior to merger and those that we could already detect we have improved angular resolution. We suspect that MBHBs carry around them matter in the form of accretion disks. Studying how the matter behaves before, during, and after merger is a subject of interest. The advantages of finding more sources and with better angular resolution prior to merger can improve our means of probing this subject.

As shown by the results of our comparison between the Fisher and MCMC parameter estimations and the single detector posterior distributions, it is not safe to use a Fisher parameter estimation for a single detector analysis. Results derived from a Fisher parameter estimation of the posterior distribution should carry with them a reason as why the Fisher parameter estimation is well-suited to tackle the problem or else proceed warily.

The waveform models used in this analysis are somewhat dated and more flexible models exist today. Future work in this area would include more advanced waveform models that include multiple harmonics as well as analysing the affects of additional detectors.

\section{Acknowledgements}
We appreciate the support of the NASA LISA foundation Science Grant 80NSSC19K0320.

\bibliography{Biblio.bib}


\end{document}